\documentclass[12pt]{article}
\usepackage[T1,T5]{fontenc}
\usepackage[latin1]{inputenc} 
\usepackage{graphicx}
\usepackage{sidecap}
\usepackage{wrapfig} 
\usepackage{subfig}
\usepackage{float}
\usepackage{amsmath} 
\usepackage{amsfonts} 
\usepackage[mathscr]{eucal}
\begin{document}  

\title{Exploring center vortices in $SU(2)$ and $SU(3)$ relativistic Yang--Mills--Higgs models}
\author{~L.~E.~Oxman$^a$, D. Vercauteren$^{bc}$ 
\\ \\
$^a$ Instituto de F\'{\i}sica, Universidade Federal Fluminense,\\
Campus da Praia Vermelha, Niter\'oi, 24210-340, RJ, Brazil.\\ \\
$^b$ Departamento de F\'{\i }sica Te\'{o}rica, Instituto de F\'{\i }sica,\\ Universidade do Estado do Rio de Janeiro,\\
Rua S\~{a}o Francisco Xavier 524, 20550-013 Maracan\~{a}, RJ, Brazil.
\\ \\ 
$^c$ Duy T\^an University, Institute of Research and Development, \\ P809, K7/25  Quang Trung, {\fontencoding{T5}\selectfont H\h ai Ch\^au, \DJ\`a N\~\abreve ng}, Vietnam.}

\maketitle     
    
\begin{abstract}        

We develop numerical tools and apply them to solve the relativistic Yang--Mills--Higgs equations in a model where the $SU(N)$ symmetry is spontaneously broken to its center.  In $SU(2)$ and $SU(3)$, we obtain the different field profiles for infinite and finite center vortices, with endpoints at external monopole sources. Exploration of parameter space permits the detection of a region where the equations get Abelianized. Finally, a general parametrization of the color structure of $SU(2)$ fields leads us to a reference point where an Abelian-like BPS bound is reconciled with $N$-ality.  
    
\end{abstract}     
      
{\bf Keywords}:  \\   

{\bf Pacs}: 11.27.+d, 11.15.-q \\       
  
\section{Introduction}

Over the past years, many lattice studies have been oriented towards obtaining the static potential from the Wilson loop average in pure Yang--Mills (YM) theories, for quarks in different representations. Asymptotic linearity \cite{Bali}, string-like behavior \cite{LW}, and $N$-ality at asymptotic distances \cite{dFK} are among the observed properties. This third one refers to the fact that string tensions depend on how the center $Z(N)$ is realized in a given $SU(N)$ quark representation. 
 
Based on the idea of dual superconductivity \cite{N}-\cite{3}, these properties have been explored by means of lattice calculations and effective models in a Higgs phase.  In the former, the possibility to capture the path-integral measure by quantum ensembles of magnetic configurations is analyzed (see \cite{ref4}-\cite{ref19}, and refs. therein). In the latter, 
phenomenological dimensionful scales are introduced from the beginning, proposing a dual superconductor
where the confining string is a smooth vortex solution to the classical equations of motion.    
This is a magnetic object in the dual description, that is supposed to effectively represent the chromoelectric confining string.
In this context, gauge models based on the $SU(N) \to Z(N)$ spontaneous symmetry breaking (SSB) pattern are attractive  (see \cite{Baker-90}-\cite{notes}, and refs. therein). In this case, the manifold of vacua is the coset ${\cal M}= SU(N)/Z(N) = Ad(SU(N))$ (the adjoint representation of $SU(N)$),
whose first homotopy group is $\Pi_1 ({\cal M}) = Z(N) $. Then, the confining string would be represented by a smooth {\it center} vortex, thus naturally leading to $N$-ality. Furthermore, as noted in ref. \cite{conf-qg}, the lack of {\it isolated} monopoles, due to the trivial $\Pi_2$ of the compact group $Ad(SU(N))$, can be interpreted as the absence of gluons in asymptotic states. However, monopoles may interpolate center vortices with different fundamental weights to form hybrid mesons, that is, a colourless quark/valence gluon/antiquark confined state.

The detailed knowledge we have about  interquark lattice potentials, for different groups and representations, makes us  wonder what the natural dual superconductor could be. With this idea in mind, the initial objective of this work is to look for and test appropriate numerical methods to solve the center vortex field equations.
These tools will permit,  in a forthcoming work, to contrast different proposals with existing lattice data, obtained from Monte Carlo simulations. A part of the data could be used to adjust the parameters, and then we could make predictions to be compared with other data. 
This type of analysis has already been considered in refs. \cite{Baker-fit-95}-\cite{su2-Michael}. In refs. \cite{Su-90}-\cite{Ingel}, an Abelian Higgs model  that essentially describes a condensate of Abelian monopoles was analyzed. For example, in ref. \cite{Ingel},
the internal structure of the flux tube, within Abelian-projected $SU(2)$ lattice gauge theory, sets the system
in the borderline between type I and type II superconductors. However, an Abelian description cannot explain $N$-ality, nor related properties such as the lattice $k$-string tensions, when $N \geq 4$ \cite{Biagio}. In the case of $SU(3)$, the fitted parameters in an effective dual QCD model also led to a similar limiting behavior (see refs. \cite{Baker-90} and \cite{Baker-Dosch}). 

In the second part of this work, we show that there exists a choice of parameters, in the model we proposed in ref. \cite{conf-qg}, where the center vortex field profiles  for $N=2,3$ satisfy Nielsen--Olesen equations, thus conciliating Abelian-like behavior with $N$-ality. 
Furthermore, in the $SU(2)$ case, a BPS bound is obtained, showing the fundamental BPS vortex is a minimum with respect to any, possibly non Abelian, field deformation. This special point in parameter space will certainly serve as a place to start exploring the model, and verify its suitableness to accommodate the various lattice data.
 
BPS bounds in a non Abelian context were previously obtained in the bosonic sector of ${\cal N}=2$ supersymmetric theories. For the embedding of 
$U(1)$-vortices, a $U(N)$ gauge field coupled to one adjoint and $N$ fundamental scalar fields  was considered \cite{Dav}. For relativistic center vortices, an appropriate limit in the parameters was needed for the BPS equations to be compatible with those of the original theory \cite{Marco}. 

\section{The Yang--Mills--Higgs model}

In order to drive $SU(N) \to Z(N)$ SSB, at least $N$ adjoint Higgs fields $\psi_I$, $I=1, \dots, d$, $d\geq N$, are required \cite{Fidel2}-\cite{HV}. In this regard, a detailed analysis of the relation between center vortex charges and magnetic weights, for different groups and representations, was carried out in ref. \cite{Konishi-Spanu}. 
  
A natural class of dual models was proposed in ref. \cite{conf-qg}. They depend on a (dual) gauge field $A_\mu$ and a set of $SU(N)$ adjoint Higgs fields,  $\psi_I \in \mathfrak{su}(N)$,
\begin{equation}
{\cal L} = \frac{1}{2} \langle D_\mu \psi_I , D^\mu \psi_I\rangle +
\frac{1}{4} \langle F_{\mu \nu}, F^{\mu \nu}\rangle - V_{\rm Higgs}(\psi_I)  \;,
\label{model}
\end{equation}
\[
D_\mu=\partial_\mu + g A_\mu \wedge\makebox[.5in]{,}
F_{\mu \nu}= \partial_\mu A_\nu -\partial_\nu A_\mu + g 
A_\mu \wedge  A_\nu  \;, 
\]
where $I$ is a flavour index, we defined $X \wedge Y = -i[X,Y]$, and the internal product is $\langle X,Y\rangle=Tr\left(Ad(X)^{\dagger}Ad(Y)\right)$, with $Ad(\cdot)$ a linear map into the adjoint representation.
The Higgs potential   
is constructed with the natural $SU(N)$ invariant terms, up to quartic order,
\[
\langle \psi_I,\psi_J \rangle 
\makebox[.3in]{,} \langle\psi_I,\psi_J \wedge \psi_K\rangle
\makebox[.3in]{,}
\langle \psi_I\wedge \psi_J,\psi_K \wedge \psi_L\rangle
\makebox[.3in]{,}
\langle \psi_I,\psi_J \rangle \langle \psi_K,\psi_L \rangle \;.
\] 
Within this class, we introduced a flavour symmetric model. 
To motivate that construction, we recall that followed for a single $SU(2)$ adjoint   Higgs field $\psi$ undergoing $SU(2)\to U(1) $ SSB. This pattern is obtained from a Higgs potential whose vacua are points on $S^2$,
\[
\langle \psi , \psi \rangle - v^2 =0 \;.
\]
A natural Higgs potential, with up to quartic terms, is then obtained by squaring the vacuum condition,
\[
V_{\rm Higgs}=  \frac{\lambda}{4} \left( \langle \psi , \psi \rangle- v^2\right)^2 \;.
\] 
Now, to get a flavour symmetric model with $SU(N) \to Z(N)$ SSB, we take $d=N^2-1$, so that the range of the flavour index coincides with that of colour. Replacing
 $I \to A=1,\dots, N^2-1$, we denote the Higgs fields as 
$\psi_A$ and initially propose a Higgs potential whose vacua satisfy,
\[
\psi_A \wedge \psi_B -  v\, f_{ABC}\, \psi_C =0\;,
\]
where $f_{ABC}$ are structure constants of $\mathfrak{su}(N)$. The vacua are given by a trivial point 
$\psi_A =0$ plus a manifold of nontrivial vacua, where  $\psi_A$ form a Lie basis. 
Of course, the space of vacua is invariant under the adjoint action of $SU(N)$ gauge transformations. 
In addition, a given Lie basis is invariant
under this action iff $U\in Z(N)$. Then, a natural potential would be obtained by squaring the condition above. Using the notation $\langle X\rangle^2 = \langle X, X \rangle$,
\begin{equation}
V_{\rm Higgs} =  \frac{\lambda}{4}\,\langle \psi_{A} \wedge \psi_{B} -f_{ABC}\, v\,\psi_{C}\rangle^{2} \;.
\label{psquare}
\end{equation}
However, for this potential the trivial and nontrivial vacua are degenerate. This can be lifted by initially  expanding the squares and then introducing general couplings for the quadratic, cubic an quartic terms,
\[
V_{\rm Higgs}= c+ \mu^2\, I_2 + \kappa \, I_3 
+ \lambda\, I_4 
\;, 
\]
\begin{equation}
 I_2=\frac{1}{2}\langle \psi_A  \rangle^2 \makebox[.3in]{,}
I_3=\frac{1}{3}\, f_{ABC} \langle \psi_A, \psi_B \wedge \psi_C \rangle \makebox[.3in]{,} 
I_4=\frac{1}{4}\, \langle \psi_A \wedge \psi_B \rangle^2 \;. 
\label{terms}
\end{equation}
Besides being gauge invariant, this potential is flavour symmetric under $Ad(SU(N))$ transformations, $\psi_A \to R_{AB} \, \psi_B$ .
The constant $c$ is chosen in order for $V_{\rm Higgs}$ to be zero when the Higgs fields assume their asymptotic vacuum values. In this manner, the asymptotic  energy density of the vortex will tend to zero, and the total energy  will be finite.
 At $\mu^{2}=\frac{2}{9}\frac{\kappa^{2}}{\lambda}$ the degenerate case is reobtained, while for $\mu^2 < \frac{2}{9}\frac{\kappa^{2}}{\lambda}$ the absolute minima are only given by nontrivial vacua. For $\kappa<0$, they are,
\[
\phi_A=  v\, S T_A S^ {-1}  \makebox[.5in]{,}
v =-\frac{\kappa}{2\lambda}\pm \sqrt{\left(\frac{\kappa}{2\lambda}\right)^2-\frac{\mu^2}{\lambda }}\;,
\label{vvalue}
\]
which verify,
\begin{equation}
\mu^2 v +\kappa v^2 + \lambda v^3 =0\;.
\label{rel-v}
\end{equation}

\subsection{The vortex between a monopole-antimonopole pair}

We can consider a finite center vortex ending at an external monopole-antimonopole pair,  which in the dual model represents a quark and an antiquark.
The pair is on the $x^1$-axis, with charges $\vec{\beta}$ and $- \vec{\beta}$,\footnote{The magnetic weight $\vec{\beta}$ is given by ($N$ times) the $(N-1)$-tuple of eigenvalues corresponding to one common eigenvector of the Cartan generators.} and placed at $x^1=-L/2$ and $x^1=+L/2$. 
In the presence of external monopoles, the energy functional is,
\begin{equation}
E= \int d^{3} x\, \left( \rho_B +
\rho_K +
V_{\rm Higgs} \right)\;, 
\end{equation}
\begin{equation}
\rho_B = \frac{1}{4} \langle F_{i j}-J_{ij}\rangle^2 
\makebox[.5in]{,} 
\rho_K=\frac{1}{2} \langle D_i \psi_A \rangle^2 \;,
\end{equation}
where $J_{ij}$ represents a pair of Dirac strings, placed on the $x^1$-axis, between the monopole locations and infinity. The energy minimization gives,
\begin{subequations} \begin{gather}
 D_j (F_{ij}-J_{ij}) = ig\, [\psi_A,D_i\psi_A]\;,
\label{YMF1} \\
 D_i D_i \psi_A = \mu^2 \psi_A + \kappa\, f_{ABC}\, \psi_B \wedge \psi_C +\lambda\, \psi_B \wedge (\psi_A \wedge \psi_B)\;.
\label{YMP1}
\end{gather} \end{subequations}

Let us consider a center vortex, ending at external monopole-like sources, with fundamental weight $\vec{\beta}$.
Because of cylindrical symmetry, all field profile functions in our ansatz are required to be $\varphi$-independent. 
The gauge field ansatz is,
\begin{equation}
A_i = \frac{1}{g} a \, \partial_i \varphi\, \vec{\beta} \cdot \vec{T}  \;,
\label{gfa}
\end{equation}
For $SU(2)$ and $SU(3)$, the weights are one and two-component tuples; they can be chosen as $\vec{\beta}=\sqrt{2}$ and $\vec{\beta}=(\sqrt{3}/2,1)$, respectively. For the Higgs fields,  
taking,
\begin{equation}
S= e^{i\varphi\, \vec{\beta} \cdot \vec{T}}\;,
\end{equation} 
and using that, for the Cartan directions, $ST_q S^{-1}= T_q$,
while, for the off-diagonal ones,
\begin{eqnarray}
S T_{\alpha} S^{-1} &=& \cos (\vec{\alpha}\cdot \vec{\beta})
\varphi \, T_{\alpha} -
\sin (\vec{\alpha}\cdot \vec{\beta}) \varphi \, T_{\bar{\alpha}} \;,\nonumber
\\
S T_{\bar{\alpha}} S^{-1} & = & \sin (\vec{\alpha}\cdot
\vec{\beta}) \varphi \, T_{\alpha} + 
\cos (\vec{\alpha}\cdot \vec{\beta}) \varphi \, T_{\bar{\alpha}} \;,
\label{rotation}
\end{eqnarray} 
we propose the form,
\begin{itemize}

\item $SU(2)$:
\begin{equation}
\psi_1 =h_{1}\, T_{1}
\makebox[.5in]{,} 
\psi_{\alpha_1}= h\, S T_{\alpha_1} S^{-1}
\makebox[.5in]{,}
\psi_{\bar{\alpha}_1}= h\, S T_{\bar{\alpha}_1} S^{-1}
\;,
\end{equation}

\item $SU(3)$:
\begin{equation}
\psi_q= h_{qp} T_p \makebox[.5in]{,} h_{qp}= \frac{1}{4} h_1\, \vec{\beta}|_q \vec{\beta}|_p + 3 h_2\,
\vec{\alpha}_2|_q \vec{\alpha}_2|_p\;,
\end{equation}\begin{equation}
\psi_{\alpha_1}= h\, S T_{\alpha_1} S^{-1}
\makebox[.5in]{,} 
\psi_{\alpha_2}=  h_0\, T_{\alpha_{2}} 
\makebox[.5in]{,}
\psi_{\alpha_3}= h\, S T_{\alpha_3} S^{-1}
\;,
\end{equation}
\begin{equation}
\psi_{\bar{\alpha}_1}= h\, S T_{\bar{\alpha}_1} S^{-1}
\makebox[.5in]{,}
\psi_{\bar{\alpha}_2}=  h_0\, T_{\bar{\alpha}_{2}} 
\makebox[.5in]{,}
\psi_{\bar{\alpha}_3}= h\, S T_{\bar{\alpha}_3} S^{-1}
\;,
\end{equation}

\end{itemize}
Note that for $SU(2)$ there is a single positive root $\alpha_1=1/\sqrt{2}$, so that the pair $\psi_{\alpha_1}$, $\psi_{\bar{\alpha}_1}$ rotates once when we go around the center vortex. On the other hand, in $SU(3)$, the three positive roots satisfy
$\vec{\alpha}_1\cdot \vec{\beta}=1$, $\vec{\alpha}_2\cdot \vec{\beta}=0$, $\vec{\alpha}_3\cdot \vec{\beta}=1$.
Then, in this case there is a pair $\psi_{\alpha_2}$, $\psi_{\bar{\alpha}_2}$ that do not rotate, while the others rotate once. In both cases, finite energy solutions require the asymptotic boundary conditions, 
\begin{equation}
a \to 1 
\makebox[.5in]{,}
 h\to v 
 \makebox[.5in]{,}
 h_1 \to v \;.
\end{equation}
(in $SU(3)$, we also have, $h_0 \to v$, $ h_2 \to v$).
There are also regularity conditions to be satisfied. The field strength tensor is,
\begin{equation}
F_{ij} 
= \frac{1}{g}\, \left\{ (\partial_i a\, \partial_j \varphi -\partial_j a\, \partial_i \varphi )  +  a [\partial_i , \partial_j] \varphi  \right\} \, \vec{\beta} \cdot \vec{T}  \;,
\label{Fij}
\end{equation}   
where $[\partial_2 ,\partial_3] \varphi =2\pi\, \delta^{(2)}(x^2,x^3)$. Then, when approaching the $x^1$-axis,
we require, 
\begin{equation}
a \to 0
\makebox[.5in]{,}
h \to 0
\makebox[.5in]{,}
 {\rm when}~ |x^1| < L/2 \;,
\end{equation}
\begin{equation}
a \to 1
\makebox[.5in]{,}
h \to v 
\makebox[.5in]{,}
 {\rm when}~ |x^1| > L/2 \;.  
\end{equation}
In this manner, the delta singularity in eq. (\ref{Fij}), present for $|x^1| > L/2 $, is cancelled against the Dirac string $J_{ij}$, leaving an energy density contribution  $(1/4)(F_{ij}-J_{ij})^2 $ that is smooth everywhere. On the other hand, the profile functions $h_0$, $h_1$ and $h_2$, associated with Higgs fields that do not rotate, are finite on the $x^1$-axis. They are not required to vanish there.

Now, let us consider curvilinear coordinates $\xi^1, \xi^2,\xi^3$ in $R^3$, 
$x=x(\xi)$. To represent vectors $ A = A^i\, \mathbf{e}_i = A_i\,  \mathbf{e}^i $, we can use either covariant or contravariant 
basis vectors $\mathbf{e}_i$ or $\mathbf{e}^i$, with Cartesian components, $\mathbf{e}_i|_{x^j} = \frac{\partial x^j}{\partial\xi^i}$, $\mathbf{e}^i|_{x^j}  = \frac{\partial \xi^i}{\partial x^j} $. The metric for contravariant and covariant coordinates satisfy $g_{ij}\, g^{jk} = \delta_i^{\;k}$,
\[dx_k \, dx_k = g_{ij} \, d\xi^i d\xi^j \makebox[.5in]{,} 
\frac{\partial \psi}{\partial x^k} \frac{\partial \psi}{\partial x^k}  =  
g^{ij} \frac{\partial \psi}{\partial \xi^i}   \frac{\partial \psi}{\partial \xi^j} \;.\]

In curvilinear coordinates, the total energy is \footnote{In these equations, the indices $i,j,\dots$, refer to curvilinear coordinates. },
\begin{eqnarray}
E &=&\int d^{3} \xi\, \sqrt{g}  \left( \frac{1}{4} \langle F_{i j}, F^{i j}\rangle + \frac{1}{2}\langle D_i \psi_A, D^i \psi_A \rangle  + V_{\rm Higgs}\right) \nonumber \\ 
&=& \int d^{3} \xi\, \sqrt{g}  \left( \frac{1}{4} g^{ik} g^{jl}\langle F_{i j}, F_{k l}\rangle + \frac{1}{2}g^{ij}\langle D_i \psi_A, D_j \psi_A \rangle  + V_{\rm Higgs}\right) \;, \nonumber 
\end{eqnarray}
while the components of the chromomagnetic field  $B = B_i\, \mathbf{e}^i $ are,
\begin{eqnarray} 
B_i 
& = &\frac{1}{2}\, g_{ij}\, \left[ \det g^{rs}\right]^{\,\frac{1}{2}} \epsilon_{jkl}\, F_{kl}  \makebox[.5in]{,} F_{kl}=  \left(  \frac{\partial A_l}{\partial \xi^k}    
-  \frac{\partial A_k}{\partial \xi^l} + g\, A_k \wedge A_l \right)  \;.
\end{eqnarray}

Let us consider any system of orthogonal coordinates, where $\xi^3$ is the polar angle with respect to the $x^1$-axis, $ \xi^3= \varphi  \in [0,2\pi)$. That is, the gauge field ansatz (\ref{gfa})  is,
\begin{equation}
A = \frac{1}{g} a \, \mathbf{e}^3\, \vec{\beta} \cdot \vec{T} \makebox[.5in]{,} a=a(\xi^1,\xi^2) \;,
\end{equation}
and the gauge field covariant components are,  
\begin{equation}
A_1 = 0  \makebox[.5in]{,} A_2=0 \makebox[.5in]{,} A_3 =\frac{a}{g}  \, \vec{\beta} \cdot \vec{T}  \;.
\end{equation}
Using the scale factors $s_i =|\mathbf{e}_i|$, and the properties,
\begin{equation}
 |\mathbf{e}^i|= s_i^{-1}
 \makebox[.5in]{,}
 g_{ii}= s^2_i \makebox[.5in]{,} g^{ii}= s_i^{-2} \makebox[.5in]{,} \sqrt{g}=s_1 s_2 s_3 \;,
\end{equation}
we obtain,
\begin{equation} 
B_1 
 =    \frac{s_1}{gs_2 s_3}\, \partial_2 a\, \vec{\beta} \cdot \vec{T} \makebox[.5in]{,}
B_2 
 =    -\frac{s_2}{g s_1 s_3}\, \partial_1 a \, \vec{\beta} \cdot \vec{T} \makebox[.5in]{,} B_3=0\;,
\end{equation}
\begin{equation}
\rho_B 
 =\frac{(N-1)}{g^2(s_3)^{2}} \left[ \left(s_2 \right)^{-2} \, (\partial_2 a)^2 + \left(s_1 \right)^{-2} \, (\partial_1 a)^2\right]\;. 
 \label{rhoB}
\end{equation} 
In addition, the covariant components of the curl of $B$ are, $\nabla \times B|_1 =\nabla \times B|_2   
 =   0 $,
\begin{equation} 
\nabla \times B|_3
 =   -\frac{s_3}{gs_1 s_2}\, \left( \partial_1 \left( \frac{s_2}{ s_1 s_3}\, \partial_1 a  \right)  
+  \partial_2 \left(  \frac{s_1}{s_2 s_3}\, \partial_2 a\right)   \right) \, \vec{\beta} \cdot \vec{T} \;.
\end{equation} 
Using our ansatz, it is easy to see that the right-hand side of eq. (\ref{YMF1}) is also along the
$\mathbf{e}^3$ direction, and that after putting $h_{\alpha}=h_{\bar{\alpha}}$ the Lie algebra directions on the left and right-hand side of the equations also coincide. In both cases, $N=2, 3$, we get,
\begin{eqnarray} 
  -\frac{s_3}{gs_1 s_2}\, \left( \partial_1 \left( \frac{s_2}{ s_1 s_3}\, \partial_1 a  \right)  
+  \partial_2 \left(  \frac{s_1}{s_2 s_3}\, \partial_2 a\right)   \right)  =g\, (1-a) h^2\;. 
\label{eqa}
\end{eqnarray} 
As $h_{\alpha}=h_{\bar{\alpha}}$, after working out
the algebra, the field equations for $\psi_\alpha$ and $\psi_{\bar{\alpha}}$ give the same information. They can be simplified using,
\[
\nabla \cdot  A \propto  \partial^2 \varphi =0 \makebox[.5in]{,} \nabla h(\xi^1, \xi^2) \cdot \nabla \varphi  = 0  \;.
\]
In what follows, we detail the remaining equations and information related with the kinetic and potential energy densities for the Higgs fields. Defining, 
\[
\hat{O} =  \partial^2 h   -\frac{(1-a)^2}{(s_ 3)^2} \,h  \makebox[.5in]{,}
\partial^2 f = \frac{1}{s_1 s_2 s_3}
\left[
\frac{\partial}{\partial \xi^1} \left( \frac{s_2 s_3}{s_1} \frac{\partial f}{\partial \xi^1} \right) +
\frac{\partial}{\partial \xi^2} \left( \frac{s_3 s_1}{s_2} \frac{\partial f}{\partial \xi^2} \right) 
\right] \;,
\]  
\noindent
$\mathbf{SU(2)}$:
\begin{subequations} \begin{gather}
\hat{O} \,h = \mu^2 h + \kappa h h_{1} +
(\lambda/2)  h(h^ 2+h^2_{1})
\label{eqh}
\\
\partial^2 h_1  = \mu^2 h_{1} + (\kappa +\lambda h_{1}) h^ 2\;.
\label{eqh1}
\\
\rho_K  =  \left[ (s_1^{-1} \partial_1 h )^2 + (s_2^{-1} \partial_2 h )^2\right] + \left( s_3^{-1}h (a-1) \right)^2   +\frac{1}{2}  \left[ (s_1^{-1} \partial_1 h_1 )^2 + (s_2^{-1} \partial_2 h_1 )^2 \right] 
 \;,
 \label{rhoK}
\\
I_2= h_1^2/2 + h^2 \makebox[.5in]{,}
I_3=  h_1  h^2 \makebox[.5in]{,}
I_4= h_1^2 h^2/2 +  h^4/4  \;, 
\\
c= -[(3/2) \mu^2 v^2 +  \kappa v^3 +(3/4) \lambda v^4 ] \;. 
\end{gather} \end{subequations}
\noindent
$\mathbf{SU(3)}$:
\begin{subequations} \begin{gather}
\hat{O} \,h =  \mu^2 h+(\kappa/6) h\, (2 h_0 +  3h_1 +  h_2 )+ (\lambda/12) h  (  6\, h^2 +2h_0^2+
 3h_1^2+  h_2^2 )\;.
 \label{eqh3}
\\
\begin{aligned}
\partial^2 h_0 
=& \mu^2 h_0  + (\kappa/3) \,( 2h_0 h_2 +h^2) + (\lambda/3)\,h_0  (h^2_0 + h^2+
 h_2^2)  \;, \\
 \partial^2 h_1   =&  \mu^2 h_1 +    \kappa h^2 + \lambda  h^2 \,h_1 \;, \\
 \partial^2 h_2   =&  \mu^2 h_2 + (\kappa/3) (2h_0^2+h^2) + (\lambda/3) h_2 (2h_0^2+h^2)   \;. 
 \label{eqh012} 
\end{aligned} \\
\begin{aligned}
\rho_K  =&  2\, [  (s_1^{-1} \partial_1 h )^2 + (s_2^{-1} \partial_2 h )^2+ \left( s_3^{-1}h (a-1) \right)^2 ]  +  (s_1^{-1} \partial_1 h_0 )^2  \\
&  + (s_2^{-1} \partial_2 h_0 )^2+ \frac{1}{2}\, [(s_1^{-1} \partial_1 h_1 )^2 + (s_2^{-1} \partial_2 h_1 )^2 + (s_1^{-1} \partial_1 h_2 )^2 + (s_2^{-1} \partial_2 h_2 )^2 ] \;.
\end{aligned} \\ 
\begin{gathered}
I_2  = 2\,h^2 +  h_0^2 +  h_1^2/2  +  h_2^2/2  
\makebox[.5in]{,} \\
I_3 =   h_1 \, h^2 +  h_2 \, h^2/3 +  (2/3)\, [h_2\,h_0^2 + h_0\, h^2] \;,
\\
I_4 = \ h_1^2\, h^2/2+ h^2_2\, h^2/6+  h_0^2\, ( h^2_2  + h^2 )/3 + h^4/2  
+ h_0^4/6 \;,
 \\
c= - [ 4 \mu^2 v^2 + (8/3)  \kappa v^3 + 2 \lambda v^4 ]\;.
\end{gathered}
\end{gather} \end{subequations}

\section{Numerical analysis}

\subsection{Infinite vortex}
\label{ivortex}
Let us initially consider the simpler case of an infinite vortex. In this way we can gain a quick understanding of how the solutions behave under the variation of parameters, and we can also check the suitableness of the numerical methods we will use. If the quarks are infinitely far apart, the problem is invariant under translations along the $x^1$-axis. In addition, due to rotational symmetry in the $(x^2,x^3)$-plane, the problem becomes purely radial (and thus one-dimensional), strongly reducing the difficulty of the numerical energy minimization. In this case, it is natural to use cylindrical coordinates,
\begin{equation} \begin{gathered}
\xi^1 \in (-\infty, \infty) \makebox[.5in]{,} \xi^2 =\rho \in [0,\infty) \makebox[.5in]{,} \xi^3= \varphi  \in [0,2\pi) \;, \\
 x^1=\xi^1   \makebox[.5in]{,} x^2= \xi^2 \cos \xi^3  \makebox[.5in]{,} x^3 = \xi^2 \sin \xi^3 \;, \\
s_1 = 1   \makebox[.5in]{,} s_2 = 1 \makebox[.5in]{,}  s_3= \xi^2\;.
\end{gathered} \end{equation}

\subsubsection{$SU(2)$}
In this section we will focus on $SU(2)$; the case of $SU(3)$ is analogous, and will be shortly touched upon in section \ref{suthree}. For $SU(2)$, using eqs. (\ref{rhoB}) and (\ref{rhoK}), the energy density per unit  length takes the form,
\begin{equation} \label{energydensityinfinite}
	2\pi \int_0^\infty  d\rho\, \rho\left(\frac1{g^2\rho^2}\, a'(\rho)^2 +  h'(\rho)^2 + \frac12 \, h_1'(\rho)^2 + 
	\frac{1}{\rho^2} h(\rho)^2 (1-a(\rho))^2  +  V_{\mathrm{Higgs}}\right)  \;.
\end{equation}

To minimize this functional, we will use a Fourier finite elements method. In short, the procedure is as follows: Initially, we modify the problem until we have one on a finite interval, with simpler boundary conditions, such that all unknown functions go to zero at the boundaries. Next, we use a Fourier series to expand them. Cutting off this series at a finite order, we plug the ans\"atze into the energy and minimize with respect to the Fourier coefficients.

With this in mind, let us start by looking at the boundary conditions. When $\rho \to0$, we must have $a \to 0$, $h \to 0 $, and $h_1$ regular. In this limit, assuming the leading term of $h$ is proportional to $\rho^i$, the equation for $h$ yields at lowest order in $\rho$,
\[
i(i - 1) + i - 1 = 0 \Leftrightarrow i = \pm1.
\]
The solution $i=+1$ satisfies the boundary conditions. Proceeding similarly with the equation for $a$, one finds that the lowest order must be either $\rho^4$ (in which case the leading terms originating from $a$ will cancel the one from the $h^2$-term), or $\rho^2$ (in which case the leading terms from $a$ cancel amongst each other  and are of lower order than that of $h^2$). We can simply assume that the Taylor series starts with a term in $\rho^2$, as this will cover both cases. Finally, no condition on the leading-order term of $h_1$ is obtained, but it turns out that the term linear in $\rho$ must vanish. We thus find the following small $\rho$ behaviors:
\begin{subequations} \label{series} \begin{gather}
a(\rho) \approx a_2\, \rho^2 + a_3\, \rho^3 + \cdots \;, \\ 
h(\rho) \approx b_1\, \rho + b_2\, \rho^2 + \cdots \;, \\ 
h_1(\rho) \approx c_0 + c_2\, \rho^2 + \cdots 
\label{hum}\;,
\end{gather} \end{subequations}
where the dots simply continue the Taylor expansions.

When $\rho \to \infty$, we need $a \to 1$, $h \to v$, and $h_1 \to v$. As the theory is massive, due to symmetry breaking, we can expect the functions to reach their asymptotic values exponentially fast. In particular, the asymptotic behavior of $a$ is found to be
\[
a(\rho) \sim 1+ \gamma \, e^{-gv\rho} \;, 
\]
with  an undetermined constant $\gamma$. Therefore, let us introduce the variable $t =
\tanh gv\rho $ with $t \in [0, 1]$. In this new variable, $a$ will be linear when $\rho\to\infty$ (and thus $t\to1$). In general, the functions $h$ and $h_1$ will not have the same exponential factor at infinity, but this will not cause any problems as long as the correct asymptotic value is reached: their behavior will simply be nonlinear in $t$. It is easily seen that the small-$\rho$ behavior in (\ref{series}) will still be valid with the replacement of $\rho$ by $t$. In effect, we have the series, 
\[
t = gv\rho + \frac{(gv\rho)^3}{3} + \dots \;,
\] 
with a vanishing second-order term, so the leading terms will remain leading, and the linear term in $h_1$ will still be absent.

Recapitulating, we can propose the ans\"atze
\begin{subequations} \begin{gather}
a(t) = t^2 + t \alpha(t) \;, \\
h(t) = vt + \eta(t) \;, \\
h_1(t) = vt^2 +  \frac{h_1(0)}2 (1 + \cos \pi t) + t \eta_1(t) \;,
\end{gather} \end{subequations}
where $\alpha(t)$, $\eta(t)$, and $\eta_1(t)$ are smooth functions in the interval $t\in [0, 1]$, that vanish at both $t=0$ and at $t=1$. Then, these new functions can be represented by means of a Fourier series only in terms of $\sin n\pi t$, $n\in\mathbb N \backslash \{0\} $.
In other words, the most general profiles can be expanded as,
\begin{subequations} \begin{gather}
a(t) = t^2 + t \sum_{n=1}^\infty a_n \sin n\pi t \;, \\
h(t) = vt + \sum_{n=1}^\infty b_n \sin n\pi t \;, \\
h_1(t) = vt^2 +  \frac{c_0}{2}\, (1 + \cos \pi t) + t \sum_{n=1}^\infty c_n \sin n\pi t \;.
\end{gather} \end{subequations}
Then, approximate static stable solutions are found by limiting the previous expressions to some finite order, plugging the functions into the energy density, and minimizing with respect to the unknown coefficients. These steps can be easily performed using the computer algebra package {\sc Mathematica}. We would like to underline that it is not necessary to transform the energy integral \eqref{energydensityinfinite} to the new variable $t$, as we can simply plug $t = \tanh gv\rho $ into our ans\"atze and do the computations using $\rho$ as a variable.
 
\begin{figure}[t]
\centering
	\subfloat[$a(\rho)$]{\includegraphics[scale=0.87]{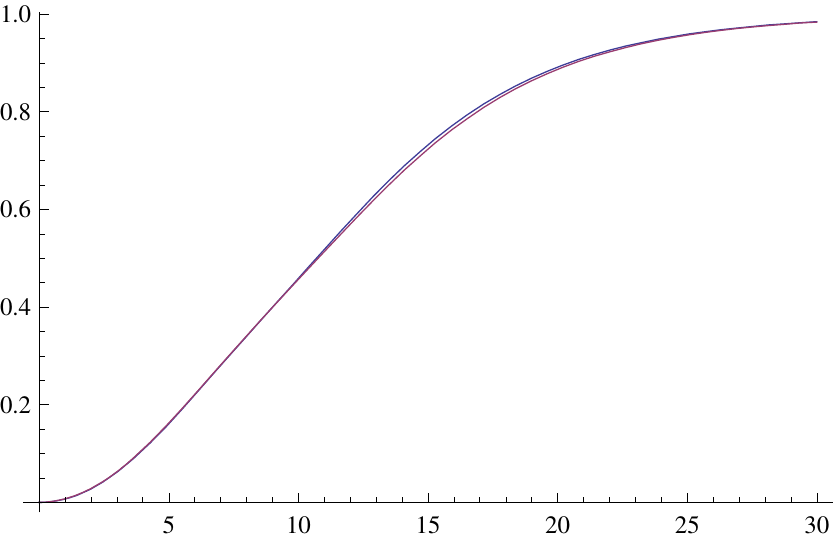}} \hfill
	\subfloat[$h(\rho)$]{\includegraphics[scale=0.87]{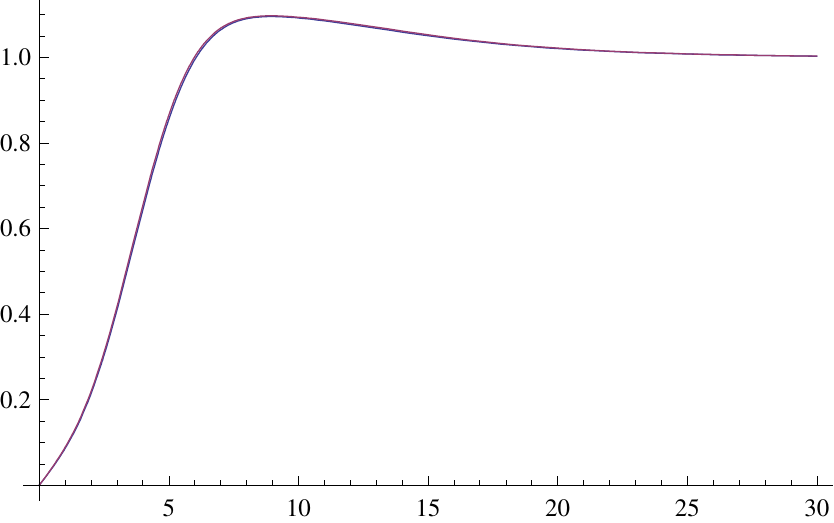}}
	\hspace{0mm} 
	\subfloat[$h_1(\rho)$]{\includegraphics[scale=0.87]{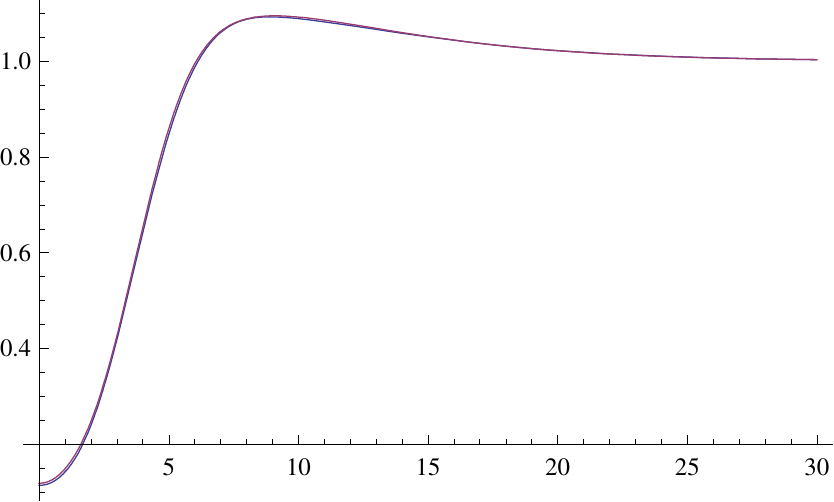}}
\caption{($SU(2)$, $g=0.1$, $\mu=1$, $\kappa=-2$, $\lambda=1$, $v=1$). The plots show the infinite vortex profiles. The purple lines include Fourier coefficients up to the fourth order (the term with $\sin4\pi t$), while the blue ones include one Fourier mode less. One can see that the numerics have converged quite well and the two successive approximations are almost on top of each other.}
\label{succ2}
\end{figure}

Figure \ref{succ2} shows the results of the numerical minimization for a certain set of parameter values. The purple lines include four Fourier modes for each of the unknown functions. For reference, the approximation with one Fourier mode less, for each profile function, is shown in blue. One can see that the approximation is already quite good and the two curves only differ noticeably in the case of $a$.

\begin{figure}
\centering
	\subfloat[$a(\rho)$]{\includegraphics[width=.47\textwidth]{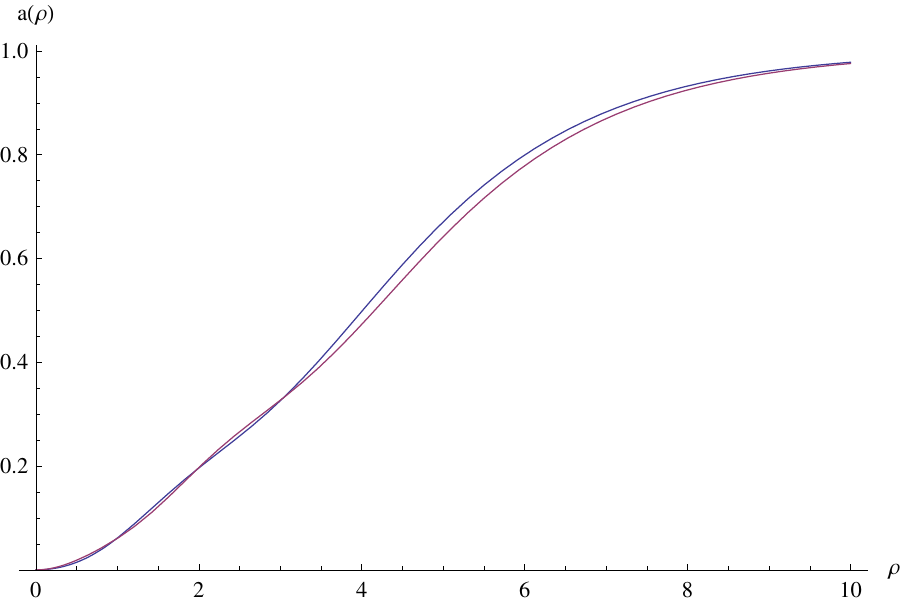}} \hfill
	\subfloat[$h(\rho)$]{\includegraphics[width=.47\textwidth]{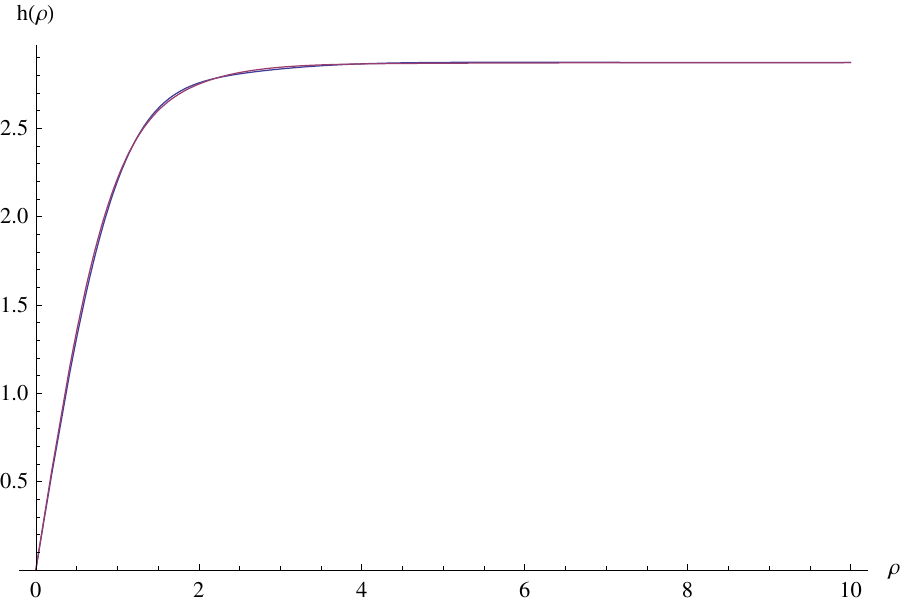}}
	\hspace{0mm}
	\subfloat[$h_0(\rho)$]{\includegraphics[width=.47\textwidth]{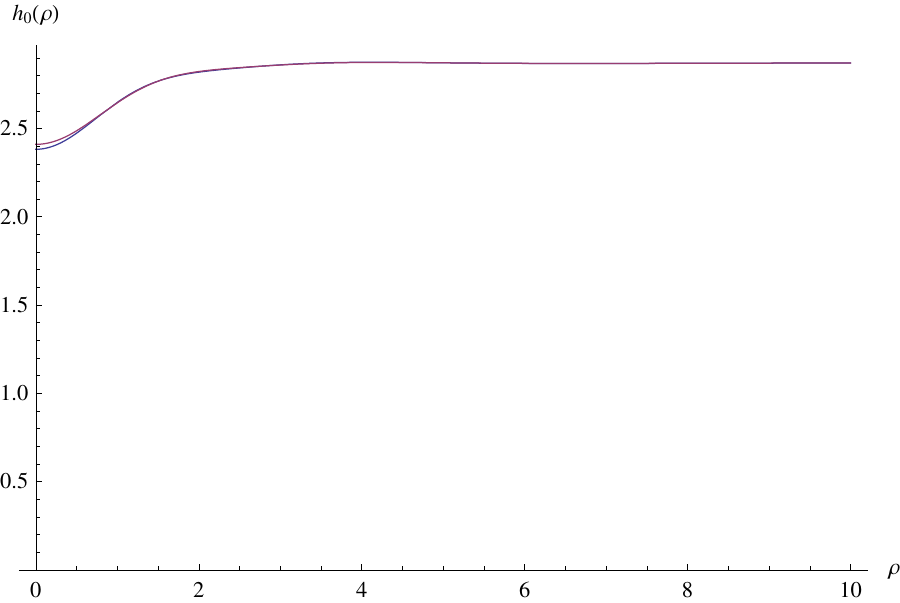}} \hfill
	\subfloat[$h_1(\rho)$]{\includegraphics[width=.47\textwidth]{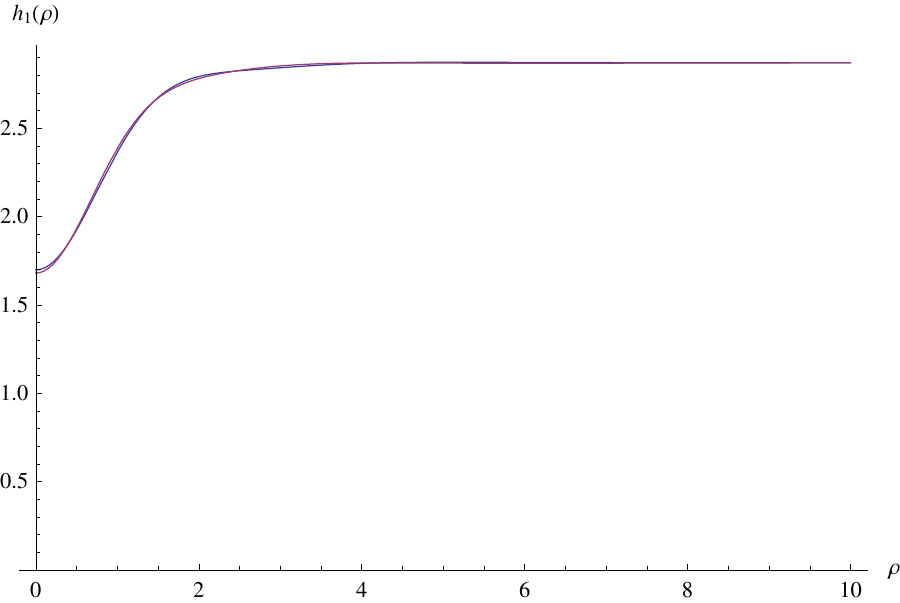}}
	\hspace{0mm}
	\subfloat[$h_2(\rho)$]{\includegraphics[width=.47\textwidth]{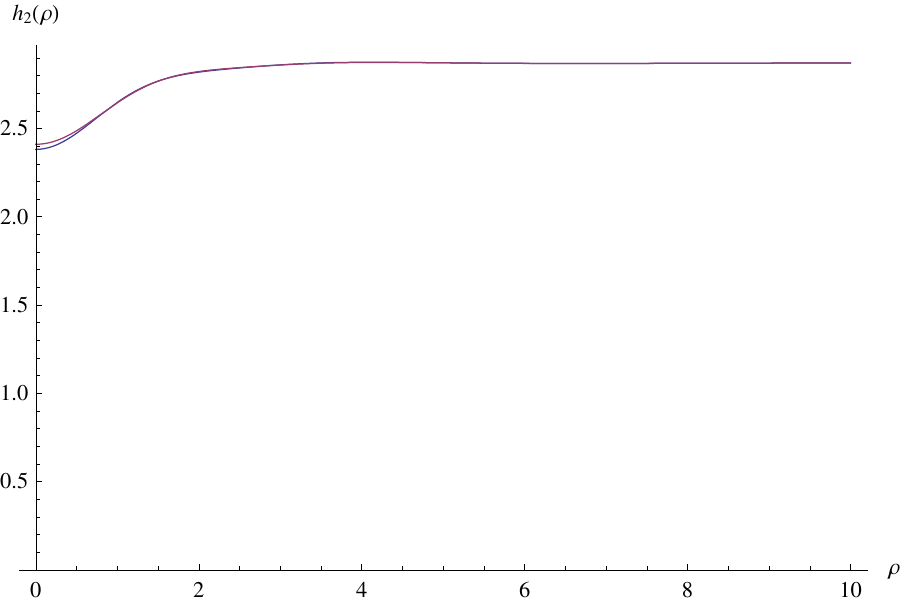}}
\caption{($SU(3)$, $g=0.1$, $\mu=1.8$, $\kappa=-4$, $\lambda=1$, $v=2.87$) The plots show, in that order, the infinite vortex profiles $a$, $h$, $h_0$, $h_1$, and $h_2$, as functions of $\rho$. Again, the purple lines include four Fourier modes for each of the unknown functions, while the blue ones only include three.}
\label{succ3} 
\end{figure}

\subsubsection{$SU(3)$} \label{suthree}
As mentioned before, the $SU(2)$ and  $SU(3)$ cases are completely analogous. The main difference is that there are two more unknown profiles, $h_0$ and $h_2$ (cf. eqs. (\ref{eqh3}) and (\ref{eqh012})), they behave exactly like $h_1$ in eq. (\ref{hum}). Figure \ref{succ3} shows our results for $SU(3)$; again, convergence is quite good and only in the case of $a$ do the approximations with three or four Fourier modes noticeable differ from each other.

\subsection{Finite vortex}

For finite vortices, it is convenient to use
prolate spheroidal coordinates with foci on the $x^1$-axis,
\begin{eqnarray} 
x^1 & = & \frac L2 \, \cosh \xi^1\, \cos \xi^2 \nonumber \\
x^2 & = & \frac L2 \, \sinh \xi^1\, \sin \xi^2 \, \cos \xi^3\nonumber \\
x^3 & = & \frac L2 \, \sinh \xi^1\, \sin \xi^2 \, \sin \xi^3 \nonumber \;,
\end{eqnarray}
where $\xi^1 $ is nonnegative, $\xi^2  \in [0,\pi]$, $\xi^3 \in [0, 2\pi)$. The scale factors are,
\begin{equation}
s_1 = s_2  = s = \frac L2 \sqrt{\sinh^2 (\xi^1) + \sin^2 (\xi^2) }
\makebox[.5in]{,}
s_3 = \frac L2 \sinh\xi^1 \sin \xi^2\;. 
\end{equation}
Defining,
$\sigma = \sinh \xi^1 \in [0,\infty)$, together with $\xi^2 = \nu$, and  $\xi^3=\varphi$, the quark and anti-quark are located at the foci described by $\sigma =0$, $\nu =0$ and $\sigma=0$, $\nu  =\pi$, respectively. The line joining them is given by $\sigma=0$, $\nu \in [0,\pi]$. The semi-infinite lines extending from $x^1=L/2$ to $+\infty$ and from $x^1=-L/2$ to $-\infty$, on the $x^1$-axis,  are given by
$\sigma \in (0, \infty)$, $\nu =0$ and $\sigma \in (0, \infty)$, $\nu  =\pi$, respectively. 

Then, the asymptotic boundary conditions, when $\sigma \to \infty$, are as follows, 
\begin{equation}
a(\sigma,\nu) \to 1
\makebox[.5in]{,}h(\sigma,\nu) \to v
\makebox[.5in]{,}
h_1 (\sigma,\nu) \to v
  \;.
\end{equation}
The regularity conditions,
when $\sigma \to 0$, $\nu \in [0,\pi]$,  are
\begin{equation}
a(\sigma, \nu) \to 0
\makebox[.5in]{,}
h(\sigma , \nu) \to 0
  \;, 
\end{equation}
while for $\nu \to 0$ or $\nu \to \pi$, with $\sigma \in (0, \infty)$, we require,
\begin{equation}
a(\sigma, \nu) \to 1
\makebox[.5in]{,}
h(\sigma , \nu) \to v
  \;.
\end{equation}
Finally, for $\sigma \in [0,\infty)$, $h_1$ is finite. 
The $SU(2)$ equations (\ref{eqa})-(\ref{eqh1}) become, 
\begin{subequations} \begin{gather}
	(1+\sigma^2) \partial_\sigma^2 a + \partial_\nu^2 a - \left(\frac{\partial_\sigma a}\sigma + \frac{\partial_\nu a}{\tan\nu}\right) = \frac{L^2}4 g^2 (\sigma^2+\sin^2\nu) h^2 (a-1) \;, \\
	(1+\sigma^2) \partial_\sigma^2 h + \partial_\nu^2 h + (1+2\sigma^2) \frac{\partial_\sigma h}\sigma + \frac{\partial_\nu h}{\tan\nu} - \frac{\sigma^2+\sin^2\nu}{\sigma^2\sin^2\nu} h (a-1)^2 \hspace{20mm} \nonumber \\
	\hspace{20mm} = \frac{L^2}4 (\sigma^2+\sin^2\nu) \left( \mu^2 h + \kappa h h_{1} + (\lambda/2)\, h(h^ 2+h^2_{1})\right) \;, \\
	(1+\sigma^2) \partial_\sigma^2 h_1 + \partial_\nu^2 h_1 + (1+2\sigma^2) \frac{\partial_\sigma h_1}\sigma + \frac{\partial_\nu h_1}{\tan\nu} \hspace{20mm} \nonumber \\
	\hspace{20mm} = \frac{L^2}{4}(\sigma^2+\sin^2\nu) \left( \mu^2 h_{1} + (\kappa +\lambda h_{1}) h^ 2\right)   \;.
\end{gather} \end{subequations}
Now, expanding around $\sigma=0$ with $\nu$-dependent coefficients, and around $\nu=0$ with $\sigma$-dependent coefficients, we obtain,
\begin{eqnarray} 
	a(\sigma,\nu) \approx a_2(\nu)\,\sigma^2 +  a_3(\nu)\,\sigma^3 + \cdots &\makebox[.5in]{,}& a(\sigma,\nu) \approx  1 + a_2(\sigma)\, \nu^2 + \cdots  \nonumber \\
	h(\sigma,\nu) \approx b_1(\nu)\, \sigma +  b_2(\nu)\, \sigma^2 + \cdots &\makebox[.5in]{,}& h(\sigma,\nu) \approx v + b_2(\sigma)\, \nu^2 + \cdots  \nonumber \\
	h_1(\sigma,\nu) \approx c_0(\nu) + c_2(\nu)\, \sigma^2 + \cdots  &\makebox[.5in]{,}& h_1(\sigma,\nu)  \approx c_0(\sigma) + c_2(\sigma)\, \nu^2 + \cdots \;, \nonumber
 \end{eqnarray}
 and similar expressions around $\nu =\pi$, in powers of $(\nu -\pi)$.
Around the quarks, the situation is more subtle. Setting $\sigma = r\cos\alpha$, $\nu = r\sin\alpha $ and considering series expansions around $r=0$, with $\alpha$-dependent coefficients,
we get,
\begin{eqnarray} 
	a(r,\alpha) & \approx & \cos^2\alpha \left[1-r^2\sin^2\alpha\, \left(\frac{1}{6}\sin^2\alpha+\frac{1}{2}\cos^2\alpha \right) + a_4(\alpha) \, r^4 + \cdots  \right] \;,	\nonumber \\ 
	h(r,\alpha) &\approx& b_0(\alpha) + \cdots \;, \nonumber \\
	h_1(r,\alpha) &\approx& c_0(\alpha) + \cdots  \;. \nonumber
 \end{eqnarray}

The following functions obey all these regularity and boundary conditions:
\begin{subequations} \begin{eqnarray} 
	a(\sigma,\nu) &\approx& \frac{\sigma^2}{\sigma^2+\sin^2\nu} \left(1-\frac{1}{2}\sin^2\nu\right) + \mathcal O(\sigma^2\nu^2) \;, \label{beha} \\
	h(\sigma,\nu) & \approx&  \frac{v\,\sigma^2}{\sigma^2+\sin^2\nu} + \mathcal O(\sigma\nu^2) \;, \\
	h_1(\sigma,\nu) & \approx& c(\sigma,\nu) \makebox[.6in]{,} \frac{\partial c}{\partial\sigma}(0,\nu) =  0 
	\makebox[.3in]{,} \frac{\partial c }{\partial\nu}(\sigma,0) = 0 
	\label{h1cond} \;.
 \end{eqnarray} \end{subequations}
 
Note that the first term in eq. (\ref{beha}) can be rewritten as,
\[
 \frac{\sigma^2}{\sigma^2+\sin^2\nu} \left(1-\frac{1}{2}\sin^2\nu\right) = \frac{z-L/2+\sqrt{(z-L/2)^2+\rho^2}}{2\sqrt{(z-L/2)^2+\rho^2}} + \, \mathcal O(({\rm distance~ to~  quark})^4)
\] 
for the quark at $x_1=+L/2$, and a similar expression, with $z-L/2$ replaced by $z+L/2$, for the  quark at $x_1=-L/2$. The corresponding contributions to the energy are,
\begin{equation} \label{selfenergy}
2\pi\int_0^\infty d\rho \int_{-\infty}^{+\infty} dz\, \frac1{4g^2} \frac\rho{((z\mp L/2)^2+\rho^2)^2} \;.
\end{equation}
These are $L$-independent divergences, which obviously correspond to the quark self-energies. As usual, we will subtract them  to get a finite total energy.

Now, we can follow a procedure similar to that used for infinite vortices.  Initially, we observe that, when $\sigma\to\infty$, $a$ behaves like $\sim 1+\gamma(\nu)\, e^{-Lgv\sigma/2}$. Then, introducing the variable $t=\tanh Lgv\sigma/2$, and defining,
\[
f(\sigma, \nu)=  \frac{\sinh^2\frac{Lgv\sigma}2-\frac{1}{3}\tanh^4\frac{Lgv\sigma}2}{\sinh^2\frac{Lgv\sigma}2+(\frac{Lgv}2)^2\sin^2\nu-\frac{1}{3}\tanh^4\frac{Lgv\sigma}2}   \makebox[.3in]{,}   g(\sigma, \nu)=\left(1-\frac{\sin^2\nu}{2 \cosh^2\frac{Lgv\sigma}2}\right)\;,
\]
we can introduce the ans\"atze,
\begin{subequations} \begin{gather}
	a(t,\nu) = f(t,\nu) g(t,\nu) + t \sin \nu \; \alpha(t,\nu) \;, \\
	h(t,\nu) = v f(t,\nu) + \sin \nu \; \eta(t,\nu) \;.
\end{gather} \end{subequations} 
The new unknown functions $\alpha(t,\nu)$ and $\eta(t,\nu)$ are smooth in the region $(\sigma, \nu) \in   [0, \infty) \times [0,\pi] $, and must vanish when either $\sigma = 0$, $\sigma \to \infty$, $\nu = 0$, or $\nu = \pi$. In terms of the variables $(t, \nu)$, they vanish on the border of the square $ [0, 1) \times [0,\pi] $, so they can be Fourier expanded in terms of the basis elements $\sin(n\pi t) \, \sin m\nu $, that is,
\begin{eqnarray} 
	a(t,\nu) &= & f(t,\nu) \, g(t,\nu)  + t\sin\nu \sum_{n,m=1}^\infty a_{nm} \sin(n\pi t) \, \sin m\nu \;, \nonumber 
	\\ 	h(t,\nu) & = & v\, f(t,\nu) + \sin\nu \sum_{n,m=1}^\infty b_{nm} \sin(n\pi t)\, \sin m\nu \;.
\end{eqnarray} 
Similarly, $h_1(t,\nu) -v$ vanishes when $t = 1$, and is finite when either $t=0$, $\nu = 0$, or $\nu = \pi$,
so it can be expanded in the basis ($p$, $m$ $\in \mathbb{Z}$),
\[
\sin \left[\frac{p\pi}{2} (t+1)\right]\, \sin m \nu \makebox[.5in]{,} \sin \left[\frac{p\pi}{2} (t+1)\right]\, \cos m \nu \;,
\]
however, the conditions in (\ref{h1cond}) select the latter basis elements, with $p = (2n+ 1)$, $n\in \mathbb{Z}$. That is, we can expand,
\[	h_1(t,\nu)  =  v + c_0\, k(t,\nu) + \sum_{n,m=0}^\infty c_{nm} \cos \Big( (n+ \tfrac12)\,\pi t \Big) \cos m\nu \;,
\]
where we defined,
\begin{equation}
	k(\sigma, \nu)= \frac{(\frac{Lgv}2)^2 \sin^2\nu}{\sinh^2\frac{Lgv\sigma}2 + (\frac{Lgv}2)^2\sin^2\nu - \frac13\tanh^4\frac{Lgv\sigma}2} \;,
\end{equation}
which is one between the quarks, thus allowing to shift the value of $h_1$ there, and zero on the other three edges. We note that terms with even $m$, in the expansions for $a$ and $h$, and with odd $m$, in the expansion for $h_1$, will always vanish due to reflection symmetry through the $(x_2,x_3)$-plane.

\begin{figure}
\centering
\subfloat[$a$]{
  \includegraphics[width=.47\textwidth]{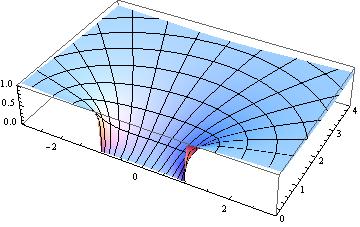}} \hfill
\subfloat[$h$]{
  \includegraphics[width=.47\textwidth]{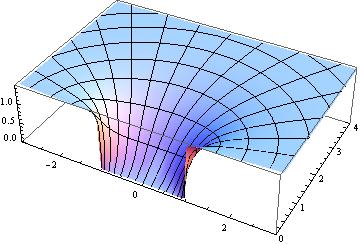}} \hfill
\hspace{0mm}
\subfloat[$h_1$]{
  \includegraphics[width=.47\textwidth]{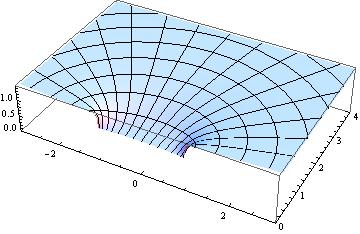}} \hfill
\subfloat[subtracted energy density]{
  \includegraphics[width=.47\textwidth]{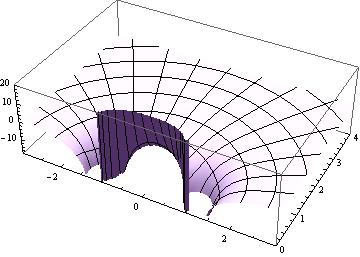}}  \hfill

\vspace{.3cm}  
\subfloat[distribution of the subtracted energy]{
  \includegraphics[width=.39\textwidth]{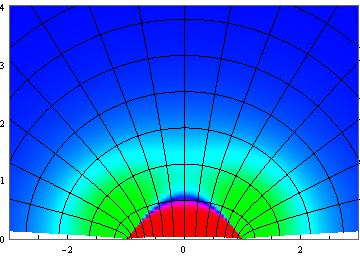}}
%  \subfloat[subtracted energy density from above]{
%  \includegraphics[width=.43\textwidth]{E-density-color+saturation-from-above.jpeg}}  
%   \subfloat[subtracted energy density from above]{
%  \includegraphics[width=.43\textwidth]{E-density-color+saturation-from-above.jpeg}}  
\hspace{0mm} 
  
\caption{($SU(2)$, $L = 2$, $g = 0.1$, $\mu = 0.9$, $\kappa = -2$, $\lambda = 1$, $v = 1.47$). Here, we show the approximation with more Fourier coefficients described in the text. The axes are $x_1$, the radial distance to the $x_1$-axis, and the different profiles. The mesh corresponds to the prolate spheroidal coordinates used in the computations. Plot (d) shows the energy density, after the quark self-energy density inside the integral in \eqref{selfenergy} has been subtracted, while the last plot shows its spatial distribution. } 
\label{fr} 
\end{figure} 

\begin{figure}[t]
\centering
\subfloat[relative error on $a$]{
  \includegraphics[width=.47\textwidth]{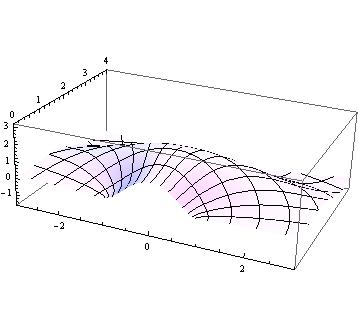}} \hfill
\subfloat[relative error on $h$]{
  \includegraphics[width=.47\textwidth]{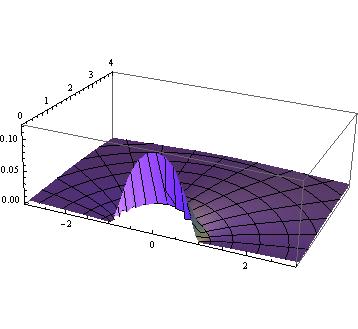}}
\hspace{0mm}
\subfloat[relative error on $h_1$]{
  \includegraphics[width=.47\textwidth]{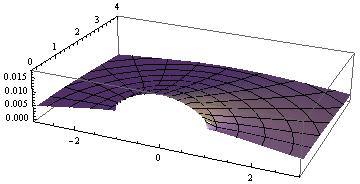}} 
\hspace{0mm}

\caption{Relative errors, in percent, for the first three profiles in Figure \ref{fr}.}
\label{percent}
\end{figure}

Finally, minimizing  with respect to the first several Fourier coefficients, we obtained the profiles showed in Fig. \ref{fr}.  There, we can see $a$, $h$, $h_1$, and the energy density, in normal Cartesian coordinates $(z,\rho)$ (the mesh lines on the plotted surfaces are the elliptic coordinates used during the computations). In this figure, we included the Fourier coefficients $a_{11}$, $a_{13}$, $a_{15}$, $a_{21}$, $a_{23}$, $a_{31}$, $b_{11}$, $b_{13}$, $b_{21}$, $c_0$, $c_{00}$, $c_{02}$, and $c_{10}$. Rerunning the minimization but only including $a_{11}$, $a_{13}$, and $a_{21}$ for $a$, keeping the previous $h$ and $h_1$ coefficients, the results are almost unchanged. Figure \ref{percent} displays (in percent) the relative errors defined as $2(a_\text{more} - a_\text{few}) / (a_\text{more} + a_\text{few})$, with $a_\text{more}$ and $a_\text{few}$ the approximation to $a$ with more, respectively fewer, Fourier coefficients (and analogously for $h$ and $h_1$).
Applying these methods, we also obtained approximate solutions in $SU(3)$. In these initial computations, we included only the $(1,1)$ mode in $a$ and $h$, and only the mode multiplying $k$ and the $(0,0)$ mode for $h_0$, $h_1$, and $h_2$. The first two profiles, $a$ and $h$, are qualitatively similar to those  obtained in $SU(2)$,
while the last three are similar to $h_1$.

%\begin{figure}
%\centering
%\subfloat[$a$]{
%  \includegraphics[width=.47\textwidth]{finitesu3a.jpeg}} \hfill
%\subfloat[$h$]{
%  \includegraphics[width=.47\textwidth]{finitesu3h.jpeg}}
%\hspace{0mm}
%\subfloat[$h_0$]{
%  \includegraphics[width=.47\textwidth]{finitesu3h0.jpeg}} \hfill
%\subfloat[$h_1$]{
%  \includegraphics[width=.47\textwidth]{finitesu3h1.jpeg}}
%\hspace{0mm}
%\subfloat[$h_2$]{
%  \includegraphics[width=.47\textwidth]{finitesu3h2.jpeg}} \hfill
%\subfloat[subtracted energy density]{
%  \includegraphics[width=.47\textwidth]{finitesu3E.jpeg}} 
%\hspace{0mm} 
%\caption{($SU(3)$, $l = 1$, $g = 0.1$, $\mu = 0.9$, $\kappa = -2$, $\lambda = 1$, $v = 1.44$).}
%\label{profsu3}
%\end{figure}  

\section{A special point in parameter space}
\label{point}

After developing appropriate numerical methods to solve the center vortex field equations, we have a tool that will permit to contrast the model with existing lattice data. For this aim, we will also need some point in parameter space to start the search for the best fit. In this respect, we recall that in ref. \cite{Su-90}, the adjusted parameters in a dual Abelian Higgs model, when fitting the lattice interquark potential, turned out to be quite close to the BPS point. Moreover, in ref. \cite{Ingel}, the internal structure of the flux tube, within Abelian-projected $SU(2)$ lattice gauge theory, was reproduced by the dual Abelian lattice description. The masses of the dual gauge and Higgs fields turned out to be quite close, again a typical property associated to a BPS point. In both works, small deviations from this point favor a weakly type-I superconductor. However, these Abelian descriptions cannot explain the observed $N$-ality of confining strings. In this section, we will show how our effective model permits to reconcile both properties at some special point in parameter space.

\begin{figure}[t]
\centering
\subfloat[$\mu = 0.8$]{
  \includegraphics[width=.47\textwidth]{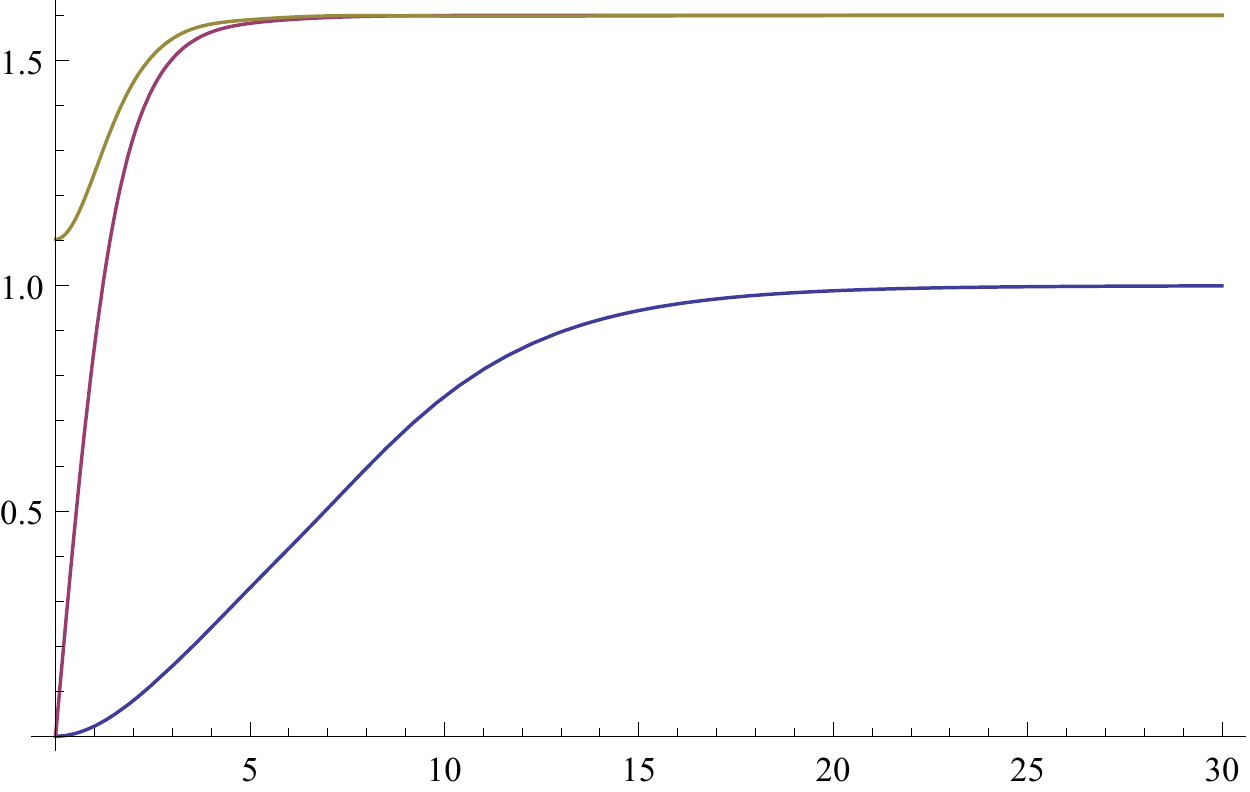}} \hfill
\subfloat[$\mu = 0.6$]{
  \includegraphics[width=.47\textwidth]{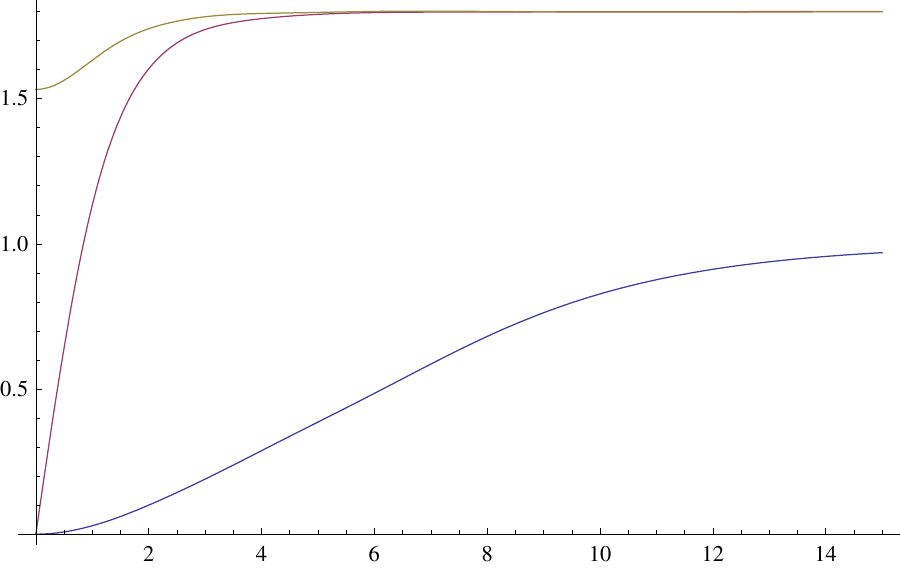}}
\hspace{0mm}
\subfloat[$\mu = 0.4$]{ 
  \includegraphics[width=.47\textwidth]{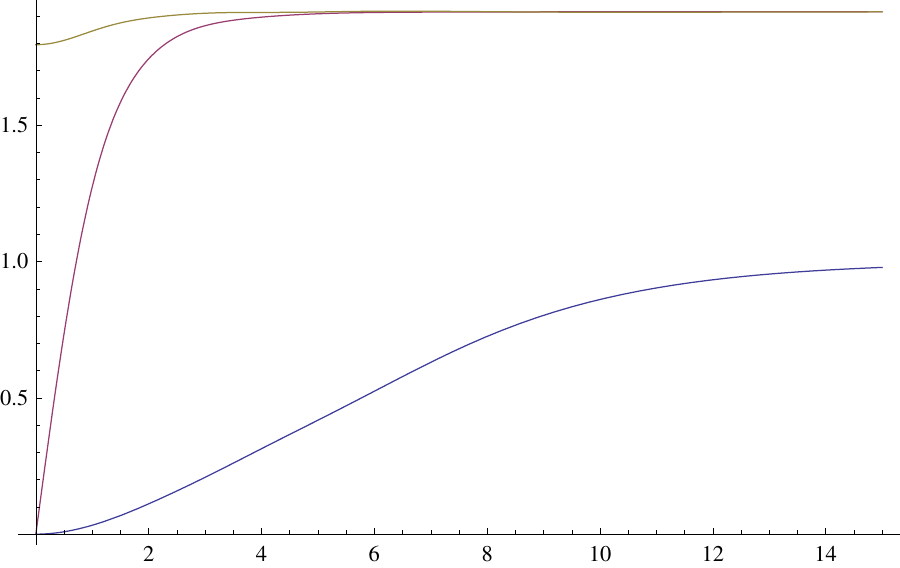}} \hfill
\subfloat[$\mu = 0.2$]{
  \includegraphics[width=.47\textwidth]{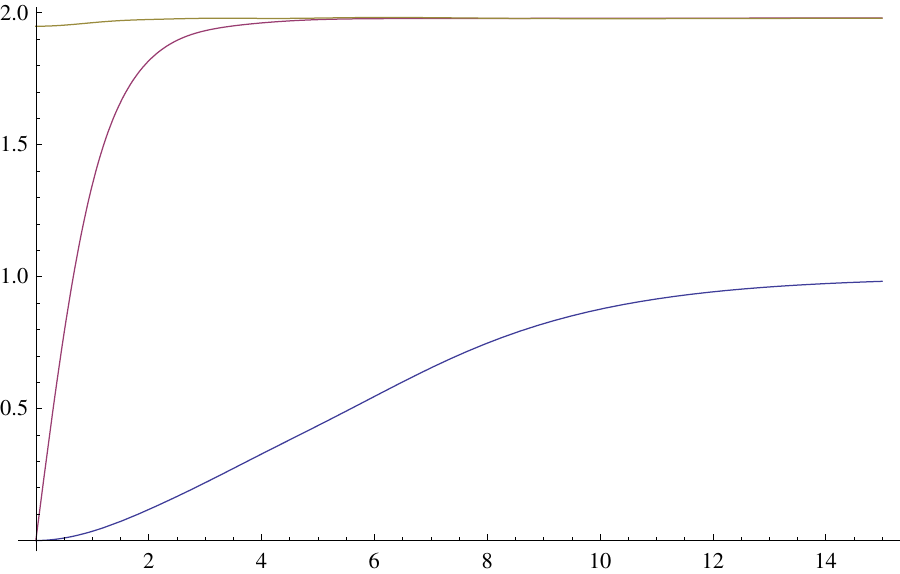}} 
\hspace{0mm}
  
\caption{($SU(2)$, $g=0.1$, $\kappa=-2$, $\lambda=1$) $a(\rho)$ (in blue), $h(\rho)$ (in red), $h_1(\rho)$ (in yellowish green).}
\label{difmu}
\end{figure}

\subsection{Center vortex BPS point}  

The $SU(2)$ center vortex profiles display a particular behavior when the mass parameter is varied. In Fig. \ref{difmu}, we display $a(\rho)$, $h(\rho)$, and $h_1(\rho)$, for $g=0.1$, $\kappa=-2$, $\lambda=1$, $v=2.87$. The four cases correspond to $\mu = 0.8, 0.6, 0.4, 0.2$, respectively. We see that $h_1(\rho)$, the profile associated with the Higgs field along the Lie algebra constant direction $T_1$,
tends to a constant function when $\mu \to 0$ (see Fig. \ref{difmu})). Indeed, at $\mu^2=0$, from eq. (\ref{rel-v}), we have $v=-\kappa/\lambda$ ($\kappa < 0$), and it can be verified, not only for $SU(2)$ but also for 
$SU(3)$, that eqs. (\ref{eqh1}) and (\ref{eqh012}) are satisfied by setting $h_1 \equiv v$ and $h_0=h_1=h_2 \equiv v$, independently of the form of $h$. That is, the field profiles for the Higgs fields that are not required to vanish on the $x^1$-axis  become frozen at the asymptotic value $v$. 
In addition, the equations for the profile $h$, namely eqs. (\ref{eqh}) and (\ref{eqh3}), both become,
\begin{equation}
 \partial^2 h   -\frac{(1-a)^2}{(s_ 3)^2} \,h =  (\lambda/2) h \, (  h^2 - v^2)  \;.
\end{equation} 
This, together with eq. (\ref{eqa}), shows that, at $\mu^2=0$, $a$ and $h$ exactly satisfy the equations for Nielsen--Olesen vortices.

As the equations for the non Abelian model get Abelianized, given that $\mu^2=0$, we may wonder whether the model (with $x^1$-translation symmetry) has a BPS point. This would permit to discuss the stability of the fundamental vortex, showing these solutions correspond to energy minima with respect to {\it any}  physical, possibly non Abelian,  change. Such BPS bound would also be useful as the Abelianized equations were obtained assuming  cylindrical symmetry, on the other hand, the bound in the non Abelian context would provide additional information. For example, at this point, there would be no forces between center vortices. A BPS bound in the bosonic sector of ${\cal N}=2$ supersymmetric theory, based on a $U(N)$ gauge field, a complex adjoint and $N$  fundamental scalars, was obtained in ref. \cite{Dav} (see also the review \cite{vortices-d}, and references therein). The BPS solutions include not only Abelian vortices embedded in the non Abelian description, but also monopoles attached to a pair of vortices.   
 
For center vortices, the initial steps we shall follow are similar to those given in ref. \cite{Oxman-BPS}, where we obtained a BPS point in a  model based on the Lagrangian (\ref{model}), (\ref{terms}),  modified by the presence of a nonrelativistic term that tends to align $\psi_1$ and $B_1$ along the same direction in the Lie algebra. In that case, we used the condition $\mu^{2}=\frac{2}{9}\frac{\kappa^{2}}{\lambda}$ (cf. eq. (\ref{terms})), where the Higgs potential becomes the perfect square in eq. (\ref{psquare}), and the decrease in energy due to the alignment term led  all BPS solutions to have zero energy. Here, we shall consider $\mu^2=0$ instead. In this case, after a general parametrization of the field color structure, the Higgs potential can also be written as a perfect square, leading to a relativistic model that accepts a BPS bound. This time, the minimum energy center vortex states will be $2\pi v^2$ and $0$, for fundamental and adjoint charges, respectively.   
 
Let us consider infinite center vortices in $SU(2)$, and the complexified variable,
\begin{equation} 
\zeta =\frac{\psi_{2}+i \psi_{3}}{\sqrt{2}}\makebox[.5in]{,}\psi_2 =\frac{\zeta+\zeta^{\dagger}}{\sqrt{2}}\makebox[.5in]{,}\psi_{3}=\frac{\zeta-\zeta^{\dagger}}{\sqrt{2}i}\;.\label{p12}
\end{equation}
In this section, we use Cartesian coordinates, and assume translation symmetry along the $x^1$-axis. Then, 
using the cyclicity property, 
\begin{equation}
\langle X,[Y,Z]\rangle=\langle[X,Z^{\dagger}],Y\rangle\;,\label{mix1}
\end{equation}
the hermiticity of $A_{i}$, and the Jacobi identity, we have ($i=2,3$), 
\begin{eqnarray}
\langle D_{i}X,D_{i}X\rangle  =\langle DX,DX\rangle+g\langle B_1,[X,X^{\dagger}]\rangle+\partial_{3}\langle X,iD_{2}X\rangle-\partial_{2}\langle X,iD_{3}X\rangle\;,\label{pI}
\end{eqnarray}
where $D=D_{2}+iD_{3}$. 
For example, taking $X=\zeta$, we obtain,
\begin{equation}
\langle D_{i}\zeta\rangle^{2}=\langle D\zeta\rangle^{2}+g\langle[\zeta,\zeta^{\dagger}],B_1\rangle+\partial_{3}\langle\zeta,iD_{2}\zeta\rangle-\partial_{2}\langle\zeta,iD_{3}\zeta\rangle\;,
\end{equation}
and the energy per unit length becomes, 
\begin{equation}
{\cal E}= \int d^2x\, \rho \makebox[.3in]{,} \rho=
\frac{1}{2} \langle D_i \psi_1\rangle^2 + \langle D\zeta\rangle^{2}+\frac{1}{2} \langle B_1\rangle^2 + g\langle[\zeta,\zeta^{\dagger}],B_1\rangle + 
V_{\rm Higgs} \;,
\end{equation}
where we have used the boundary condition,
\begin{equation}
D_{i}\zeta\to0\;.\makebox[.5in]{{\rm for}}(x^{2},x^{3})\to\infty\;,
\end{equation}
needed for a finite ${\cal E}$. 
Now, in the general parametrization $\psi_A = \Psi|_{AB}\, T_B$, the $3\times 3$ real matrix $\Psi$ can be always decomposed as the product of a lower triangular matrix $L$ times an orthogonal matrix.
If $\Psi$ is invertible, requiring the diagonal elements of $L$ to be positive, the factorization is unique. As a matrix in $O(3)$ is a sign times an $SO(3)$ matrix, and $\det L = L_{11} L_{22} L_{33}$, when $\det \Psi > 0$ ($\det \Psi  <0$) there is a unique decomposition $\Psi = +  L R^T $ ($\Psi = - L R^T $), with $R^T \in SO(3)$. That is, we can represent,
\begin{eqnarray}
\psi_1 &= & L_{11}\, n_1 \nonumber \\
 \psi_2 &= & L_{21}\, n_1 + L_{22}\, n_2   \nonumber \\
 \psi_3 &= & L_{31}\, n_1 + L_{32}\, n_2 + L_{33} \, n_3 \;,
 \label{Les}
\end{eqnarray}
where $n_A = S T_A S^{-1} = T_B\, R_{BA}$. For $SU(2)$, in our conventions, we have $f_{123}=1/\sqrt{2}$. 
Then, setting $\mu^2=0$, using the asymptotic vacuum value $v= -\kappa/\lambda$ ($\kappa < 0$), and
 $c= (1/4)\, \lambda v^4$, needed to have zero potential energy at the vacuum, the Higgs potential can be cast in the form,
\begin{eqnarray} 
V_{\rm Higgs} 
&=&  \frac{\lambda}{4}\, \left[ L_{11}^2( L_{22} - L_{33})^2 + 2\, L_{22}L_{33}\, (L_{11}-v)^2    +
(L_{22} L_{33} -v^2)^2 \right] \nonumber \\ 
&& + \frac{\lambda}{4}\, \left[ (L_{11} L_{32})^2 + (L_{21} L_{33})^2 + (L_{21}L_{32}-L_{22}L_{31})^2  \right] \;.
\end{eqnarray} 
Finally, expanding
\begin{equation}
B_1 = B_{11}\, n_1 + B_{12}\, n_2 +  B_{13}\, n_3 \;,
\end{equation}
 we note that at  $\lambda = g^2$  we can write,  
\begin{eqnarray}
\rho &=& \frac{1}{2} \langle D_i \psi_1\rangle^2 + \langle D\zeta\rangle^{2} +\frac{1}{2}\, \left[ B_{11}+ \frac{g}{\sqrt{2}}\, (L_{22} L_{33} -v^2) \right]^2 + \frac{gv^2}{\sqrt{2}}\,B_{11} 
\nonumber \\
&& +\frac{1}{2}\, \left[ B_{12}-\frac{g}{\sqrt{2}}\, L_{21} L_{33} \right]^2 +\frac{1}{2}\, \left[ B_{13}+ \frac{g}{\sqrt{2}}\, (L_{21}L_{32}-L_{22}L_{31}) \right]^2 \nonumber \\
&&  + \frac{g^2}{4}\, \left[ L_{11}^2( L_{22} - L_{33})^2 + 2\, L_{22}L_{33}\, (L_{11}-v)^2 + (L_{11} L_{32})^2  \right] \;. 
\label{bound-n1}
\end{eqnarray}
Setting the squares to zero, we get the BPS equations. Among them, $L_{11}=v$ and
$D_i \psi_1 = 0$ lead to the condition $D_i n_1 = 0$. Defining the fields (see ref. \cite{conf-qg} and references therein), 
\begin{equation}
C_i^A = -(1/g) f^{ABC} \langle n_B, \partial_i n_C\rangle \;,
\end{equation}
$A=1,2,3$, the general solution to this condition is,
\begin{equation}
A_i = a_i \, n_1 - C_i^a n_a \makebox[.5in]{,} 
\end{equation}
$a=2,3$, and the resulting magnetic field is along $n_1$, that is, $B_{12}=0$, $B_{13}=0$,
\begin{equation}
B_{11} =  \partial_2 a_3 -\partial_3 a_2-\frac{g}{2}\, \epsilon_{1jk} f^{1ab}  C^a_j C^b_k \;.
\end{equation} 
For a finite energy, the asymptotic behavior $a_i \to -C_i^1$ is required, as the combination $-C_i^A n_A $ is {\it locally} a pure gauge field. The associated field strength is,
\[
F_{ij} = -F^A_{ij}(C)\, n_A  \makebox[.5in]{,} F^A_{ij}(C) =
 \partial_i C_j^A- \partial_j C_i^A +g\,  f^{ABC}  C^B_i C^C_j \;.
\]
If $n_1$ is well defined everywhere, then the only nonzero component is $F^1_{ij}(C)$, and it is concentrated on the string where $n_2$, $n_3$ are ill-defined, as occurs at the center vortex guiding center (cf. eq. (\ref{rotation})). Noting that,
\[
\int d^2x\,( \partial_2 a_3 -\partial_3 a_2 )  = \oint  dx_i \,a_i = -\oint  dx_i \,C_i^1 = \int d^2x\,( \partial_2 C_3^1 -\partial_3 C_2^1 ) \;,
\]
we obtain,
\begin{eqnarray} 
\int d^2x\,B_{11} = -\frac{1}{2} \int d^2x\, \epsilon_{1jk} F^1_{jk}(C)   \;.
\end{eqnarray}

Then, under the condition $D_i n_1 =0$, the flux of $B_1$ projected along $n_1$ is topological and invariant under {\it regular} gauge transformations. 
For a single center vortex, this flux turns out to be $2\pi \sqrt{2}/g$ (see  \cite{conf-qg}), and the energy per unit length becomes $ {\cal E}  = 2\pi v^2$. Setting the remaining squares to zero gives, 
\[L_{22} = L_{33} = h  \makebox[.5in]{,}  L_{21} = L_{31} = L_{32}=0 \;, \]   
\begin{equation}
D\zeta=0  \makebox[.5in]{,}
B_{11} =\frac{g}{\sqrt{2}}\, \left(v^2 - h^2\right)\;.
\label{pe1}
\end{equation}
Finally, using that $A_i$ is locally given by $S(a_i + C_i^1)\,T_1S^{-1} + \frac{i}{g} S \partial_i S^{-1}$,
we get,
\begin{equation}
a_i  = \frac{\sqrt{2}}{g}\, \epsilon_{ij} \partial_j \ln h - C_i^1\makebox[.5in]{,}
-\partial^2 \ln h  + 2\pi  \delta^{(2)}(x_2,x_3) =\frac{g^2}{2}\, \left(v^2 - h^2\right)\;. 
\end{equation}

It is important to underline that the compatibility of the BPS and YMH equations is due to the fact that 
the energy density (\ref{bound-n1}) is bounded by (a constant times) the projection $\langle n_1, B_1 \rangle$, 
$\langle n_1 \rangle^2 =1$. This would not be the case if the projection were along the dynamical Higgs field $ \psi_1 $, as occurs at $\mu^{2}=\frac{2}{9}\frac{\kappa^{2}}{\lambda}$, where it is necessary to redefine the  model by subtracting a nonrelativistic term proportional to $\langle \psi_1, B_1 \rangle$ \cite{Oxman-BPS}. A similar  field-dependent projection is observed in the BPS center vortex bound of ref. \cite{Marco}. In that case, to keep the equations compatible, an appropriate limitting behaviour of the parameters was needed.

\section{Conclusions}

The detailed knowledge we have about  interquark lattice potentials makes us  wonder what the natural effective description for the Yang--Mills vacuum could be. This search can be guided by the symmetries, the way they are realized, the identification of large distance relevant terms in the functional energy and, of course, by the lattice data. Here, we analyzed a natural class of models with $SU(N) \to Z(N)$ SSB. This is an interesting SSB pattern, 
as the confining string would be represented by a smooth center vortex, thus incorporating $N$-ality. Initially, we tested a numerical method to solve the center vortex field equations, obtaining the solutions for infinite and finite center vortices, the latter running between a monopole and an antimonopole, representing the external quark and antiquark, respectively.  

In fact, the lattice interquark potential has already been adjusted in different phenomenological models, and
the lattice data seem to set the parameters close to the interface between type I and type II superconductors. For instance, this has been observed in a dual Abelian Higgs model that essentially describes a condensate of Abelian monopoles. However, we know that in that case, $N$-ality cannot be accommodated. In the second part of this work, we reconciled both properties in our framework. We showed numerically and analytically (for $SU(2)$ and $SU(3)$) that there is a region in parameter space where the field equations freeze some Higgs profiles to a constant vacuum value. In this region, the profiles for the gauge field along a local Cartan direction, and for the Higgs fields that rotate, exactly satisfy Nielsen--Olesen equations. So we can already conclude that in our non Abelian context, the fitting of lattice data will be as good as in the Abelian one, with the advantage of implementing $N$-ality.   

Moreover, in the case of $SU(2)$, after freezing one of the Higgs fields at a local vacuum value, we derived a  BPS bound that is topological and gauge invariant under regular gauge transformations. This point provides the type-I/type-II superconductor interface. The steps followed are similar to those previously given to 
derive a  BPS point, at $\mu^{2}=\frac{2}{9}\frac{\kappa^{2}}{\lambda}$, $\lambda=g^2$, by the inclusion of a nonrelativistic interaction that tends to align the magnetic field and one of the Higgs fields along the same Lie algebra direction. As a consequence, in that work, all BPS solutions had zero energy. Here, we have shown that to get a BPS bound at $\mu^{2}= 0$, $\lambda=g^2$, the model requires no alignment term. This time, the minimum energy center vortex states are  $2\pi v^2$ and $0$, for fundamental ($z=1$) and adjoint ($z=2$) charges, respectively.

These are essential tools that will permit to determine, in a forthcoming work, the appropriate model that is compatible with the various observables already computed in the lattice, as normal and hybrid  potentials, and the energy density profiles.

\section*{Acknowledgements}

We would like to thank Roman H\" ollwieser and Gabriel Santos-Rosa for useful discussions.
The Conselho Nacional de Desenvolvimento Cient\'{\i}fico e Tecnol\'{o}gico (CNPq), the Coordena\c c\~ao de Aperfei\c coamento de Pessoal de N\'{\i}vel Superior (CAPES), and FAPERJ are acknowledged for the financial support.

\end{document}